\documentclass[11pt,twoside]{book}
\usepackage{konkolyproc2}
\usepackage{longtable}
\usepackage{amsmath,amssymb}
\usepackage{graphicx}
\usepackage{lscape}
\usepackage{index}
\usepackage{natbib}
\usepackage{bigdelim}
\usepackage{multirow}
\makeindex

\begin{document}

\pagestyle{myheadings}
\setcounter{equation}{0}\setcounter{figure}{0}\setcounter{footnote}{0}\setcounter{section}{0}\setcounter{table}{0}\setcounter{page}{1}
\markboth{W.~A. Dziembowski}{RRL2015 Conf. Papers}
\title{Nonradial Oscillations in Classical Pulsating Stars. Predictions and Discoveries}
\author{Wojciech A. Dziembowski$^{1,2}$ }
\affil{$^1$Nicolaus Copernicus Astronomical Center, ul. Bartycka 18,
00-716 Warszawa,\\ $^2$Warsaw University Observatory, Aleje
Ujazdowskie 4, 00-478 Warszawa, Poland}

\begin{abstract}
After a brief historical introduction and recalling basic concepts of stellar oscillation theory, I focus my review on interpretation of secondary periodicities found in RR~Lyrae stars and Cepheids as a manifestation of nonradial mode excitation.
\end{abstract}

\section{Introduction}
Even after \cite{ledoux} seminal paper on nonradial oscillations in $\beta$ Canis Maioris, the idea that such oscillation may be spontaneously excited in stars was not generally accepted. The paradigm of radial pulsation was too deeply rooted. When I \citep{Dziembowski1971} found that in the radiative cores of evolved stars all nonradial modes become gravity waves of very short wavelength, which leads to a large energy loss, I thought that this explained  why Cepheids and RR~Lyrae stars excite radial modes only. However, I soon became sceptical about it.

Few years later, independently \cite{osaki} and I \citep{Dziembowski1977} showed that models of classical pulsators are unstable to excitation of not only radial but also certain nonradial modes beginning from some moderate degrees. The instability was found for unfitted envelope upon assuming certain boundary condition, which each of us derived in different way and there was an unessential difference in its form.

Complete stellar models are needed to study low degree modes. First survey of low degree modes in an RR star model by \cite{VDK98} revealed very dense spectra of unstable modes in the instability range of radial modes but their growth rates were much lower. The authors believed that the dipolar modes, which reach growth rate maxima very close to radial modes, may be excited and cause the Blazhko effect.

\section{Nonradial oscillation driven by the opacity mechanism}

\begin{figure}[!ht]
\centering
\includegraphics[width=.85\textwidth]{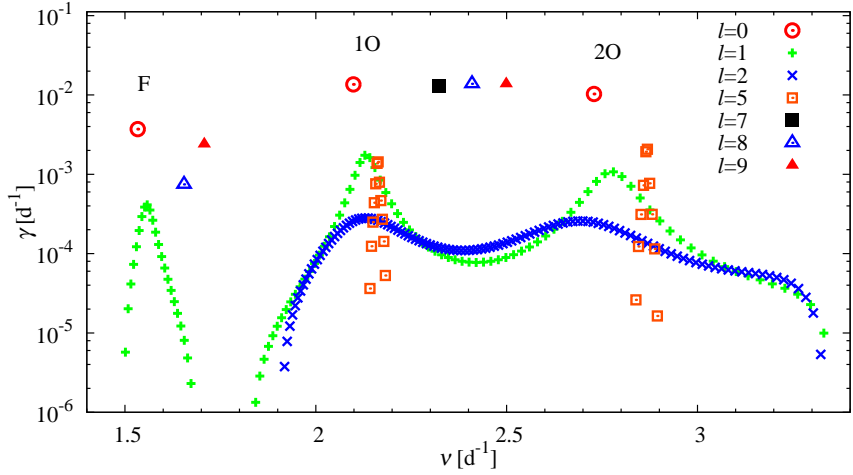}
\caption{Driving rates of unstable modes in an RR~Lyrae star model. The model was selected from the evolutionary track starting from ZAHB with the parameters $M=0.67M_\odot$, $Y_{\rm surf}=0.23$, $Z_{\rm surf}=0.001$, and $\alpha_{\rm MLT}=1.99$. The current model parameters are $X_{\rm c}=0.687$, $\log(L/L_\odot)=1.717$, $\log T_{\rm eff}=3.805$.}
\label{fig:umodes}
\end{figure}

In Fig.~\ref{fig:umodes} I show an example of the two types of unstable modes in the model of RR~Lyrae star. All modes shown in this figure owe their instability to the same effect but its efficiency, measured by the driving rate, depends on the relative amplitude in the driving zone; the zone is localized in the outer part of a star. Therefore, the high driving rates are found for modes that are best trapped in the outer (acoustic) cavity, which extends down to the place where
$$\max({\cal N},{\cal L}_\ell)=\omega.$$
I adopted here the standard notation for the angular frequency of mode, $\omega=2\pi\nu$,
for the Brunt-V\"{a}is\"{a}l\"{a} frequency
$${\cal N}=\sqrt{g\left(\frac{1}{\Gamma_1}\frac{{\rm d}\log P}{{\rm d} r}-\frac{\rm d\log\rho}{{\rm d}r}\right)},$$
and the Lamb frequency
$$ {\cal L}_\ell=\sqrt{\ell(\ell+1)}\frac{c}{r}.$$
There is no inner (gravity wave) cavity for radial modes. For nonradial modes its top and the bottom are determined by the equality
$$\min({\cal N},{\cal L}_\ell)=\omega.$$
The layer between the two cavities, called evanescent zone, determines the strength of trapping. It is the weakest at $\ell=2$ and 3 because the two critical frequencies are close to each other but further increase of $\ell$ results in stronger trapping.

At $\ell=5$, the modes have still the mixed character. We see in Fig.~\ref{fig:umodes} few unstable modes which are partially trapped in the outer cavity. Their frequencies depend on the properties of the two cavities. Already at $\ell=7$ there is only one unstable mode associated with the radial mode. For them we will use the $f,p_1,...$ notation adopted in nonradial oscillation theory. The general category of such modes was termed by \cite{VDK98} the Strongly Trapped Unstable (STU) modes.

In the gravity-wave cavity, the ${\cal N}/\omega$  ratio reaches very high values that allows to represent the eigenfunctions of the solution as a superposition of the running wave carrying the energy upward and downward. The neglect of the former leads to the boundary condition which may be safely used only for unstable modes. This is why the letter U is essential in the acronym STU.

\section{Why it may be easier to detect nonradial modes at the harmonic than at mode frequencies}

So far efforts to identify the nature of the periodicity in the $(0.6-0.65)P_{1\rm O}$ range in Cepheids and in RR Lyrae stars (hitherto called $P_{\rm x}$) were made under the assumption that they are due to individual mode periods. Here I will show that they are likely to be due to harmonics.

\begin{figure}[!ht]
\centering
\includegraphics[width=.6\textwidth]{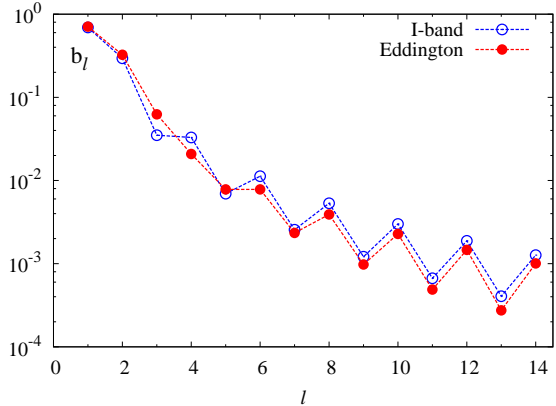}
\caption{Cancellation factor with two versions of limb darkening law.} \label{fig:bl}
\end{figure}

Let us consider oscillations in a slowly rotating star written in the inertial spherical coordinate system ($r,\,\theta,\,\phi$) centered in the star with polar axis aligned with the rotation axis. Then the relative perturbation of the surface radius caused by modes of the angular degree, $\ell$, and the azimuthal order $m\le0$ (these two numbers will be given only when needed) may be written as follows,
$$ \frac{\delta R}{R}=\epsilon_z\tilde P_\ell^0\cos(\alpha_z-\omega t)+\tilde P_\ell^m[\epsilon_p\cos(\psi_p-\omega t)+\epsilon_r\cos(\psi_r-\omega t)],$$
where the subscripts $z$, $p$ and $r$, refer to zonal, prograde and retrograde modes, respectively, $\tilde P_\ell^m$ are proportional to the usual associated Legendre functions, except that they are normalized so that each $\epsilon$ represents the rms value of $\delta R/R$. For the compactness, I defined
$$\psi_p=\alpha_p+m[\phi-(1-C)\Omega],\quad\mbox{ and }\psi_r=\alpha_r-m[\phi-(1-C)\Omega],$$
where $C$ is the Ledoux constant. $\alpha_{r,\ell}^m$ and $\epsilon_{r,\ell}^m$ denote phases and the rms amplitudes, which are arbitrary in linear theory. Then, the approximate expression for the observable photometric amplitude in the specified passband (here I) is
\begin{equation}
A_{\rm I,obs}(i)=\epsilon|b_{\ell,\rm I}\tilde P_\ell^m(\cos\theta_{\rm obs})|
 \sqrt{(B_T|f|)^2-2B_TB_R|f|\cos(\arg f)+B_R^2},
\end{equation}
where $i=\theta_{\rm obs}$,
$$b_{\ell,{\rm I}}=\int_0^1h_{\rm I} P_\ell\mu d\mu,\quad
 \mbox{with } h_{\rm I}\mbox{ being the limb darkening coefficient},$$
$$B_T=\frac{1}{4}\frac{\partial\log F_I}{\partial\log T_{\rm eff}},\quad\mbox{ and }\quad B_R=(\ell-1)(\ell+2).$$
$f$ is complex coefficient that connects the local radiative flux perturbation to the radius perturbation. The $b_\ell$ coefficient describes cancelation of contributions from various parts of the stellar disc. Its behaviour for low degree modes is plotted in Fig.~\ref{fig:bl}. We observe that even at moderate $\ell$ it drops by 2 to 3 orders of magnitude as compared to radial mode.

The most obvious and likely dominant effect contributing to peaks at the harmonic is the nonlinearity of the $\delta F(\delta R)$ dependence. These peaks arise first from the terms that are proportional to $(\delta R/R)^2$. If we square the  expression for $\delta R/R$, because of orthogonality of the associated Legendre functions, only squared functions survive the integration over $\cos\theta$.  The only $\phi$-dependent coefficients that survive the integration over the azimuth angle are constant terms  and terms with $2\omega t$ dependence.  The squares of the associated Legendre functions may be put in the form of the following series,
$$(\tilde{P}_\ell^m)^2=\sum_{k=0}^{2\ell}a_k\tilde{P}_{2k}^{2m}.$$
In view of the steep decline of $b_\ell$, it is justified to neglect all higher order terms. Then, using $a_0=2-\delta_{m,0}$, we obtain the following expression for $(\delta R/R)^2$:
\begin{equation}
\left(\frac{\delta
R}{R}\right)^2=\frac{1}{2}\left[\epsilon^2_z\cos(2\alpha_z-2\omega
t)+2\epsilon_p\epsilon_r\cos(\alpha_p+\alpha_r-2\omega t)\right].
\end{equation}
We have only harmonic time-dependence and no cancelation, which for the peaks at individual frequencies is described by the $b_\ell$
factor. This is why we may have high amplitude at the harmonic and, in addition, all components of the multiplet contribute to $2\omega$. Lets note that in the linear approximation all the components contribute at different frequencies. Therefore we can expect much simpler structure in the frequency spectrum near $2\omega$ than at $\omega$. Further, we expect much less variability connected with individual frequencies at around $2\omega$ than at $\omega$ where complex interaction within the multiplet occurs, even for the simple dipole case as shown by \cite{bgs95}.

\section{Interpretation of the $\bf P_{\rm x}$ periodicity in 1O Cepheids}

\begin{figure}[!ht]
\includegraphics[width=1.0\textwidth]{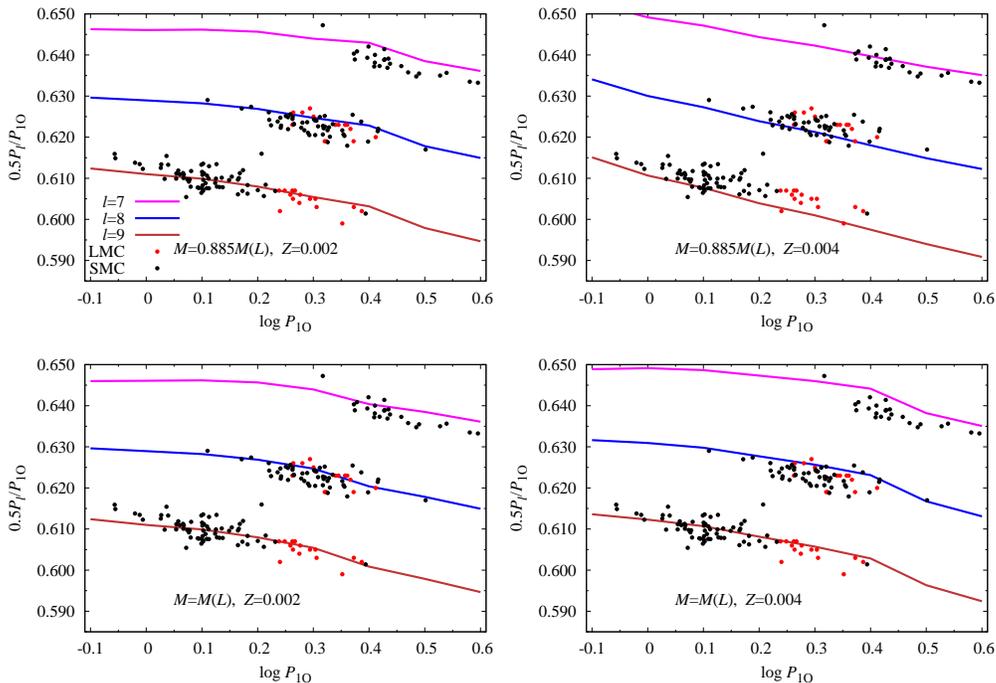}
\caption{OGLE-III data on $P_{\rm x}$ periods (black dots -- SMC, red dots -- LMC) confronted in the Petersen diagram with the values of the $0.5P_\ell/P_{\rm 1O}$ ratios calculated for stellar envelope models. The lines connect the interpolated ratios at the eight selected values of $\log P_{\rm 1O}$ within the instability strip.}\label{fig:cep}
\end{figure}

Models used for comparison with data in Fig.~\ref{fig:cep} are the same as in Dziembowski (2012). In particular, the adopted mass-luminosity relation is $\log L/L_\odot=3.05+3.6(\log M/M_\odot - 0.602).$
The observational data are also the same. The difference is in association of the three sequences in the Petersen diagram well visible in OGLE-III data for Cepheids in the Magellanic Clouds \citep{ogle_lmc,ogle_smc}, with oscillation modes. In my 2012 paper, if $P_{\rm x}$ corresponds to unstable modes then the only possibility are f-modes at $\ell=42, 46$, and 50. There is a number of difficulties for such identification. For instance, while the absence of odd degrees may be blamed to the visibility pattern shown in Fig.~\ref{fig:bl}, the conspicuous lack of $\ell=44$ and $48$ seems impossible to explain. Moreover, there is a good deal of uncertainty in claiming instability of the invoked high degree modes.  The {\it bona fide} driving operates solely in the H-ionization zone, in which convective flux dominates and calculations are notoriously not credible.

In the new picture, the top sequence is due to excitation of the $\ell=7$ in stars with $\log L/L_\odot\in[3.15,\,3.45]$ and $M/M_\odot\in[4.7,\,5.2]$, if $Z=0.002$.  The range of luminosity is similar at higher metal abundance but implied masses are lower. At $\ell=7$ instability occurs only at $\log P_{\rm 1O}\gtrsim0.4$, which corresponds to $\log L/L_\odot\gtrsim3.1$. The remaining two sequences are well accounted for with unstable $\ell=8$ and 9 modes in models satisfying adopted mass-luminosity relation at $Z=0.002$. It is of some interest that this relatively low $Z$ fits to both the SMC and LMC objects, which is well below value of 0.008 usually adopted for massive stars in the LMC. This may explain the much lower incidence rate of $P_{\rm x}$ periodicity in this galaxy. The sequence formed by the $\ell=9$ modes extends from $\log P_{\rm 1O}=-0.05$ to $0.4$, which corresponds to $\log L/L_\odot\in[2.63,\,3.3]$ and $M/M_\odot\in[3.05,\,4.7]$. Let us notice that the LMC stars occupy the long period part of the sequence.

With the new interpretation of the $P_{\rm x}$ periodicity the three sequences seen in the data may be identified with the first three STU f-modes. In the outer part of stellar envelopes, in the driving layers, these modes are similar  to radial F-modes. Thus, what we see in the 1O/X stars may be regarded as a form of double-mode pulsation, not much different from the F/1O pulsations and presumably arising in similar circumstances.

\section{Interpretation of the $\bf P_{\rm x}$ periodicity in RRc stars}

The $P_{\rm x}$ periodicity was first discovered in RRc stars. However, only after \cite{nsm1,nsm} analysis of the OGLE-IV data on RRc stars in  the Galactic bulge, we have a sufficiently large sample for studying general properties of RRc stars exhibiting the $P_{\rm x}$ periodicity.

\begin{figure}[!ht]
\includegraphics[width=1.0\textwidth]{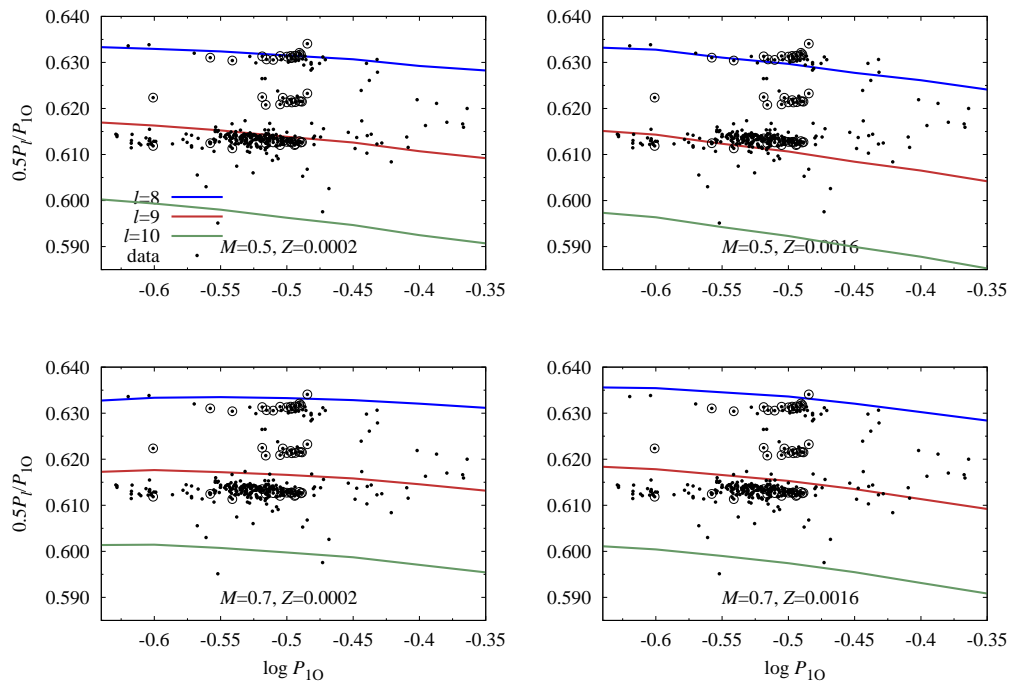}
\caption{Similar to Fig.~\ref{fig:cep} but for data for the Galactic bulge RRc stars and suitable envelope models. The encircled dots refer to multiple $P_{\rm x}$  peaks present in one object} \label{fig:rrl}
\end{figure}

In Fig.~\ref{fig:rrl}, like in Fig.~\ref{fig:cep}, we also see three sequences. Within the adopted interpretation of the $P_{\rm x}$, the bottom and top sequences may only be associated with $\ell=9$ and 8 modes, respectively. Then the intermediate sequence may be associated with the corresponding combination peaks. This is consistent with large fraction of peaks that have  partners in one or both surrounding sequences (encircled stars in Fig.~\ref{fig:rrl}).

As we may see in Fig.~\ref{fig:rrl}, the lower mass models fit to the sequences better. At the long period end, the best fit is achieved for the lower values of $Z$ and $\log(L/L_\odot)\approx1.65$. At the opposite end, where the fit is better for the higher $Z$, $\log(L/L_\odot)\approx1.55$. This is consistent with the known trend of the metallicity-luminosity dependence in RR~Lyrae stars.

\section{Problems with the long periodicity}

Recently, \cite{nsd} found in the OGLE-IV Galactic bulge photometric data 11 objects classified as RRc showing hitherto unknown periodicity, close to $1.458P_{1\rm O}$. Periods of F-modes in RR~Lyrae stars are always shorter. Thus, if this periodicity is to be associated with an oscillation mode, it is most likely stable. Possible exception to consider are dipolar modes, which as we see in Fig.~\ref{fig:umodes}, may stay unstable at periods longer than $P_{\rm F}$. However, it seems impossible to explain why the modes located far from the driving rate maximum should always be selected.

\begin{figure}[!ht]
\centering
\includegraphics[width=.75\textwidth]{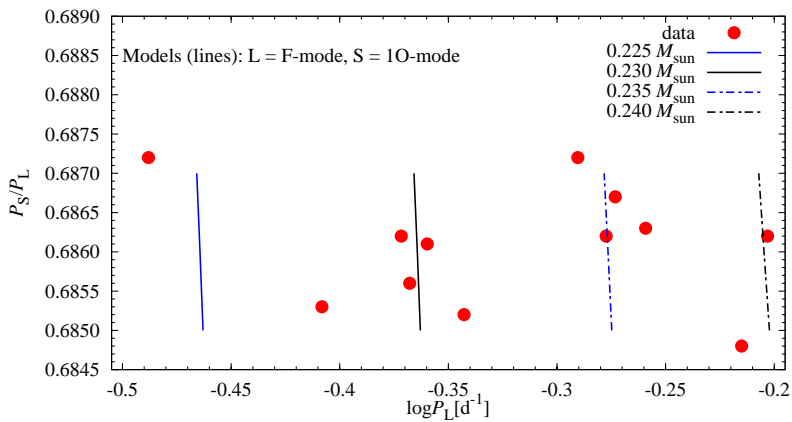}
\caption{Low-mass giant models in the Petersen diagram.} \label{fig:striped}
\end{figure}

Stellar envelope models with $1.458P_{1\rm O}=P_{\rm F}$, that is, $P_{\rm S}/P_{\rm L}\approx0.686$, may be constructed if these objects are not RRc stars but stripped giants similar to OGLE-BLG-RRLYR-02792 \citep{bep}. Such models may reproduce positions of these objects in standard Petersen diagram, as shown in Fig.~\ref{fig:striped}. This is an attractive alternative to the interpretation of the $1.458P_{1\rm O}$ periodicity in terms of nonradial modes. Unfortunately, it is not supported by data for the {\it Kepler} object KIC 9453114 where this periodicity coexists with that at $0.614P_{1\rm O}$ \citep{pamsm15}. The latter, more significant, is fully consistent with the positions occupied by the $\ell=9$ modes in the low mass RRc stars and it would be at about half distance between the $\ell=9$ and 8 (or 10) if the masses are as low as required with the stripped giant interpretation. On the other hand, interpretation of the $1.458P_{1\rm O}$ periodicity in KIC 9453114 as a consequence of nonradial mode excitation has also unacceptable implications.

\section{Conclusions}

Nearly forty years after the first results from linear nonadiabatic calculation for models of Cepheids and RR~Lyrae stars, showing that there are unstable modes in the models, we finally have a convincing evidence that such modes are indeed excited in stars of both types. The evidence is based on the interpretation of periodicity $(0.6-0.65)P_{1\rm O}$ (named $P_{\rm x}$) found in 1O Cepheids and RRc stars. In the new interpretation, this periodicity arises from harmonics of nonradial f-modes effectively trapped in the outer part of envelope and, as shown in 1977, are driven by the opacity mechanism with a similar rate as the radial F-mode. In Cepheids the modes have angular degrees from $\ell=7$ to 9, in RR~Lyrae only 8 and 9 but, in addition, the signature of the 8+9 combination peak is visible.

The signals at $2P_{\rm x}$, originally interpreted as subharmonic, are now attributed to mode frequencies. The signatures extend over a broad range, because in fact there are $2\ell+1$ components corresponding to different $m$-values. Since Cepheids and RR~Lyrae stars are slow rotators, we cannot expect the components to be resolved, as the frequency peaks are wide and variable. This variability may arise from resonant interaction between $m=0$ and the $\pm m$ pairs. Reliable central frequencies are best obtained from $P_{\rm x}$, because all components contribute to the peak at 2$\omega$ only. It remains to be seen whether they provide useful constraints on stellar global parameters.

The nature of the $1.458P_{1\rm O}$ periodicity in RRc stars is still a puzzle. It could be attributed to $P_{1\rm O}/P_{\rm F}$ period ratio if the objects are low-mass stripped giants rather than RRc stars. However, data from one star observed with {\it Kepler} contradict such interpretation, but they cannot be used to support the interpretation invoking the nonradial mode either.

 {\bf Acknowledgements.} This paper is based on materials included in part in the joint publication, coauthored by Radek Smolec, to be published soon. I would like to thank Radek Smolec and my wife Anna for their help in preparation of the manuscript. My research is supported by the Polish National Science Centre through grant DEC-2012/05/B/ST9/03932.

\end{document}